\documentclass{iopart}

\usepackage[english]{babel}
\usepackage{amssymb}
\usepackage{epsfig}
\usepackage{color}
\usepackage{subfigure}
\usepackage{graphicx}
\usepackage{color}
\usepackage{longtable}
\renewcommand{\(}{\left (}
\renewcommand{\)}{\right )}
\renewcommand{\[}{\left [}
\renewcommand{\]}{\right ]}

\begin{document}

\title[$dd$ excitations in cuprates measured by RIXS]{Energy and symmetry of $dd$ excitations in undoped layered cuprates measured by Cu $L_3$ resonant inelastic x-ray scattering}

\author{M~Moretti~Sala$^1$, V~Bisogni$^2$\footnote{Present address: IFW Dresden, Postfach 270116, D-01171 Dresden, Germany}, C~Aruta$^3$, G~Balestrino$^4$, H~Berger$^5$, N~B~Brookes$^2$, G~M~de Luca$^3$, D~Di~Castro$^4$, M~Grioni$^5$, M~Guarise$^5$, P~G~Medaglia$^4$, F~Miletto~Granozio$^3$, M~Minola$^1$, P~Perna$^3$, M~Radovic$^3$\footnote{Present address: Swiss Light Source, Paul Scherrer Institut, CH-5232 Villigen PSI, Switzerland}, M~Salluzzo$^3$, T~Schmitt$^6$, K~J~Zhou$^6$, L~Braicovich$^7$ and G~Ghiringhelli$^7$}

\address{$^1$CNISM and Dipartimento di Fisica, Politecnico di Milano, Piazza Leonardo da Vinci 32, I-20133 Milano, Italy}

\address{$^2$European Synchrotron Radiation Facility, Bo\^{\i}te Postale 220, F-38043 Grenoble, France}

\address{$^3$CNR-SPIN and Dipartimento di Scienze Fisiche, Universit$\grave{a}$ di Napoli ``Federico II'', Complesso di Monte S.~Angelo, Via Cinthia, I-80126 Napoli, Italy}

\address{$^4$CNR-SPIN and Dipartimento di Ingegneria Meccanica, Universit\'{a} di Roma Tor Vergata, Via del Politecnico 1, I-00133 Roma, Italy}

\address{$^5$Ecole Polytechnique F\'ed\'erale de Lausanne (EPFL), Institut de Physique de la Mati\'ere Condens\'ee, CH-1015 Lausanne, Switzerland}

\address{$^6$Swiss Light Source, Paul Scherrer Institut, CH-5232 Villigen PSI, Switzerland}

\address{$^7$CNR-SPIN and Dipartimento di Fisica, Politecnico di Milano, piazza Leonardo da Vinci 32, I-20133 Milano, Italy}

\ead{marco2.moretti@mail.polimi.it}

\begin{abstract}

We measured high resolution Cu $L_3$ edge resonant inelastic x-ray scattering (RIXS) of the undoped cuprates La$_2$CuO$_4$, Sr$_2$CuO$_2$Cl$_2$, CaCuO$_2$ and NdBa$_2$Cu$_3$O$_6$. The dominant spectral features were assigned to $dd$ excitations and we extensively studied their polarization and scattering geometry dependence. In a pure ionic picture, we calculated the theoretical cross sections for those excitations and used them to fit the experimental data with excellent agreement. By doing so, we were able to determine the energy and symmetry of Cu-3$d$ states for the four systems with unprecedented accuracy and confidence. The values of the effective parameters could be obtained for the single ion crystal field model but not for a simple two-dimensional cluster model. The firm experimental assessment of $dd$ excitation energies carries important consequences for the physics of high $T_c$ superconductors. On one hand, having found that the minimum energy of orbital excitation is always $\geq 1.4$ eV, i.e., well above the mid-infrared spectral range, leaves to  magnetic excitations (up to 300 meV) a major role in Cooper pairing in cuprates. On the other hand, it has become possible to study quantitatively the effective influence of $dd$ excitations on the superconducting gap in cuprates.

\end{abstract}

\submitto{\NJP}
\maketitle

\section{Introduction}
Despite the enormous effort devoted to the study of high $T_c$ superconductivity in cuprates, a general consensus on the underlying mechanisms is still lacking. It is commonly agreed that low energy elementary excitations should play a crucial role in the formation of conduction electron Cooper pairs, so that considerable efforts are being devoted to make a link between phonon and magnon spectra and high $T_c$ in cuprates \cite{SupercBook2008}. Lattice modes (phonons), which are at the basis of conventional superconductivity as explained by BCS theory, lie below 90 meV, an energy seemingly incompatible with critical temperatures peaking as high as 130 K. And the possible role of spin excitations (magnons) is still debated: the superexchange coupling in layered cuprates is exceptionally large ($J\simeq130$ meV) and gives rise to magnetic excitations up to almost 300 meV. Although in doped superconducting materials the long range magnetic order is lost, short range magnetic correlations persist \cite{Kastner1998}, and magnetic excitations survive across the whole reciprocal space as recently shown by Braicovich \etal\cite{LucioPhaseSep}. The possible role of magnetic excitations onto Cooper pairing has been the object of several works in the past \cite{Scalapino1995} and is still very actively studied at present \cite{LeTacon:unp}.

In this context orbital ($dd$) excitations, which correspond to a change in the \emph{symmetry} of the occupied Cu-$3d$ orbitals, have been attracting much less attention, because their energy is generally much higher (typically $> 1.5$~eV). However, starting from an advanced use of Eliashberg equations \cite{Holcomb1996}, Little \etal~\cite{Little2007} argued that \emph{dd} excitations are also possibly implicated in the mechanisms of high $T_c$ superconductivity. Indeed a relation between the Cu to apical-oxygen distance and $T_c$ had been found already several years ago by Ohta and coworkers~\cite{Ohta1991}, who proposed a rationale based on Madelung potentials. They summarized their results in the so-called ``Maekawa plots'', reporting the dependence of $T_c$ on the energies needed to transfer the Cu-$3d_{x^2-y^2}$ hole to a Cu-$3d_{3z^2-r^2}$ orbital or to the O-$2p$ states. More recently Sakakibara \etal\cite{Sakakibara2010} have refined the theoretical analysis by a two-orbital model applied to the model structure of La$_{2-x}$Sr$_x$CuO$_4$ and HgBa$_2$CuO$_{4+\delta}$. In their calculations the key parameter is the energy difference between the Cu-$3d_{x^2-y^2}$ and Cu-$3d_{3z^2-r^2}$ states ($\Delta E_{e_g}$): $d$-wave superconductivity is favored by a larger energy splitting because of a reduced contribution of the $d_{3z^2-r^2}$ state to Fermi surface. In the mentioned works the authors based their discussion on theoretically calculated $dd$ excitation energies rather than on experimental ones. However, there is no consensus on the actual values because theoretical and experimental estimates vary from author to author. It is thus timely to provide to the theory of superconductivity a firmer experimental basis. Here we provide all the $dd$ excitation energies in several undoped layered cuprates measured by resonant inelastic x-ray scattering (RIXS) at the Cu $L_3$ edge.

The Cu ion in 2D layered cuprates is nominally divalent (Cu$^{2+}$), corresponding to a $3d^9$ electronic configuration. It is generally accepted that in the ground state the $3d$ hole on Cu is mainly found in a $d_{x^2-y^2}$ orbital~\cite{CTChenPRL1992}. $dd$ excitations correspond to a change in the symmetry of occupied $3d$ states. In particular, for Cu$^{2+}$, the hole is excited from the $d_{x^2-y^2}$ symmetry to the $d_{3z^2-r^2}$, $d_{xy}$, $d_{xz}$ or $d_{yz}$ orbitals. $dd$ excitations are an important ingredient for the understanding of the electronic structure of cuprates beside their possible direct implication in the mechanism of high $T_c$ superconductivity itself. In analogy we can look at NiO, whose \emph{dd} excitations have been studied by optical spectroscopies~\cite{Newman1959,Fujimori1984}, electron energy loss spectroscopy (EELS)~\cite{Fromme1995,Fromme1996} and, recently, by RIXS~\cite{Ghiringhelli2005,Chiuzbaian2006,GhiringhelliPRLNiO}. However, although extensively investigated, \emph{dd} excitations in cuprates are still object of debate because of the lack of conclusive experimental results.

Most of the experimental basis comes from optical spectroscopy. In the optical absorption of La$_2$CuO$_4$ and Sr$_2$CuO$_2$Cl$_2$, Perkins \etal~\cite{PerkinsPRL1993} found a sharp feature at 0.41 (0.36) eV, which they assigned to transitions to the $d_{3z^2-r^2}$ orbital. The same authors identified the transition to the $d_{xy}$ orbital to occur at 1.50 eV in Sr$_2$CuO$_2$Cl$_2$. Electroreflectance measurements by Falck \etal~\cite{FalckPRB1994} on La$_2$CuO$_4$ revealed the transitions to the $d_{xy}$ and $d_{xz/yz}$ orbitals to be at 1.40 and 1.60 eV, respectively.  This scenario was compatible with crystal field calculations accompanying Cu $L_3$ RIXS measurements by Ghiringhelli \etal~\cite{GhiringhelliPRLdd} that located the transitions to the $d_{3z^2-r^2}$, $d_{xy}$ and the doubly degenerate $d_{xz/yz}$ orbitals at 0.41 (1.17), 1.38 (1.29) and 1.51 (1.69) eV for La$_2$CuO$_4$ (Sr$_2$CuO$_2$Cl$_2$). However, Lorenzana and Sawatzky~\cite{LorentzanaPRL1995} had previously proposed that the feature at $\sim 0.4$ eV in the optical absorption spectra is rather due to a phonon assisted bimagnon excitation. And more recent RIXS results by Braicovich \etal~\cite{BraicovichPRL2009} confirmed the latter assignment for La$_2$CuO$_4$ and CaCuO$_2$: the mid-infrared feature is due to magnetic excitations, dispersing up to $\sim0.40$ eV, and strong $dd$ excitations are found around 1.5 - 2.5 eV. This fact is also compatible with the optical Raman scattering measurements by Salomon \etal~\cite{SalamonPRB1995}, who identified the transition to the $d_{xy}$ orbital at 1.70 eV for La$_2$CuO$_4$ and 1.35 eV for Sr$_2$CuO$_2$Cl$_2$. Moreover Cu $M_{2,3}$ edge RIXS measurements in Sr$_2$CuO$_2$Cl$_2$ by Kuiper \etal~\cite{KuiperPRL1998} had already located the transitions to the $d_{3z^2-r^2}$, $d_{xy}$ and $d_{xz/yz}$ orbitals at 1.50, 1.35 and 1.70 eV, respectively (see also \cite{GhiringhelliPRLdd} for Cu $L_3$ RIXS results). Finally direct first principle calculations by Middlemiss \etal~\cite{MiddlemissJPCM2008} support these findings too.

By taking advantage of the recent experimental improvements in the field of soft x-ray RIXS we address here the problem of determining the $dd$ excitation energies in cuprates in a systematic way. We have measured Cu $L_3$ RIXS spectra of La$_2$CuO$_4$ (LCO), Sr$_2$CuO$_2$Cl$_2$ (SCOC), CaCuO$_2$ (CCO) and NdBa$_2$Cu$_3$O$_6$ (NdBCO) at several sample orientations.  \emph {dd} excitations are shown here to have local character with no or very little dispersion vs the in-plane transferred momentum \emph{$\textbf{q}_{\parallel}$}. Using single ion theoretical cross sections we could determine the $dd$ excitation energies for all samples with a high degree of confidence.

\section{Resonant inelastic x-ray scattering}

$L_{2,3}$ edge RIXS is emerging as a powerful technique for the study of neutral excitations in 3$d$ transition metal (TM) oxides and cuprates in particular. Tanaka and Kotani~\cite{Tanaka1993} suggested that \emph{dd} and charge transfer (CT) excitations in cuprates can be seen with this technique. It should be emphasized that, while forbidden in the optical absorption spectra by dipole selection rules, \emph{dd} excitations are allowed in RIXS due to two consecutive dipole transitions. In fact, the scattering process taking place in $L_3$ edge RIXS can be seen as follows. Initially a 2\emph{p$_{3/2}$} electron is resonantly transferred in the 3\emph{d} shell through the absorption of an x-ray photon: the system is then in a highly excited state with a deep core hole. Secondly the system decays via the transition of a 3\emph{d} electron into the 2\emph{p$_{3/2}$} states and the emission of a photon. As the intermediate state, characterized by a $2p$ core hole and an extra $3d$ electron, is not observed, the whole process has to be described at the second order by the Kramers-Heisenberg equation. We deal then with an energy loss spectroscopy, i.e., an inelastic scattering of x-ray photons that leaves the solid in an excited state. The energy and momentum of the final state are known from the measured variation of energy and momentum of the scattered photons. In this work we restrict to the cases where the system is left in a final configuration corresponding to a $dd$ or spin-flip excitation.

It must be noted that local spin-flip excitations are not eigenstates of the 2D antiferromagnetically ordered lattice. Rather spin waves (magnons) are excited by Cu $L_3$ RIXS. Those collective excitations are known to disperse in energy as function of their momentum and are traditionally mapped by inelastic neutron scattering (INS). In the specific case of layered cuprates magnetic excitations are particularly difficult to measure with neutrons due to their very high energy and only rather recently high quality results were obtained for La$_2$CuO$_4$ ~\cite{ColdeaPRL2001}.  Very recently the same type of measurements were made by Braicovich \etal using Cu $L_3$ RIXS~\cite{BraicovichPRL2009,LucioPhaseSep,GrioniPRLSCOC}. Thanks to the relatively large momentum carried by soft x rays, the \emph{$\textbf{q}$}-dependence of single and multiple magnons could be investigated over approximately 2/3 of the first Brillouin Zone (BZ). Those results were also supported theoretically~\cite{AmentPRL2009}. It is thus  clear that, thanks to the advances in the instrumentation~\cite{AXES,DiNardo,SAXES} that led the experimental linewidth in the 100 meV range, Cu $L_3$ edge RIXS has become a complementary technique to INS for the measurement of spin waves dispersion.

On the other hand \emph{dd} excitations have been studied by Cu $L_3$~\cite{GhiringhelliPRLdd} and $M_{2,3}$ edge (3\emph{p}-3\emph{d}) RIXS~\cite{KuiperPRL1998} with partial success. At $M$ the lower photon energy provides a potentially better energy resolution as compared to $L$ edge; however, because of the limited momentum carried by photons the portion of the BZ which can be probed is limited to a region close to the $\Gamma$ point. Any possible dispersion is thus hardly detectable. Furthermore the insufficient spin-orbit splitting of the 3$p$ states and the extreme weakness of the inelastic cross section with respect to the elastic scattering make the analysis and interpretation of the experimental data very difficult. It is interesting to note that \emph{dd} excitations have also been seen at Cu $K$ edge RIXS~\cite{EllisPRB2008} but in this case $dd$ are much weaker than CT excitations because they are indirectly excited.

\section{Experimental information}

\begin{figure}
\begin{indented}
  \item{}\includegraphics[width=10cm]{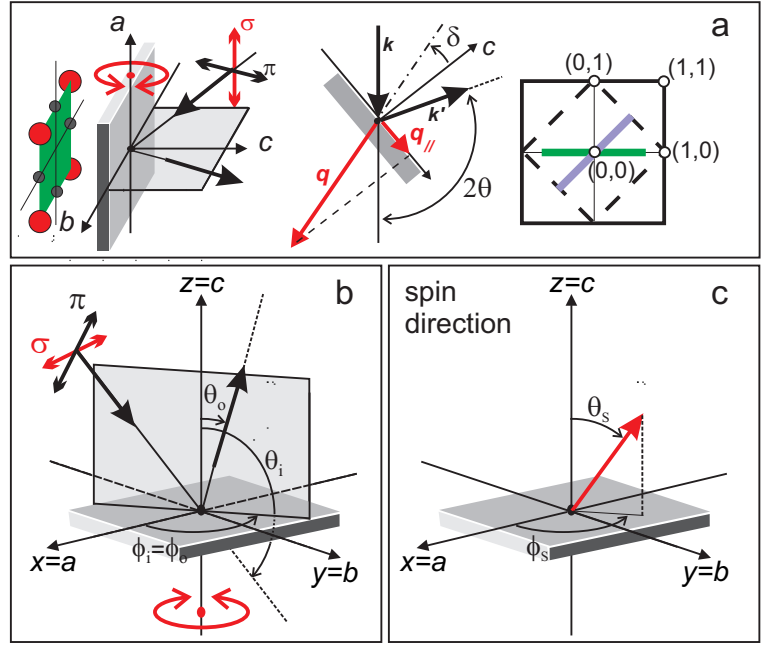}
  \caption{Color on-line. Panels a and b: The experimental geometry is shown. The incoming beam hits the sample surface (assumed to be parallel to the $ab$ plane) at incident angles $\theta_{i}$ and $\phi_{i}$ and the outgoing beam is collected at the angles $\theta_{o}$ and $\phi_{o}$. The scattering angle 2$\theta$ is fixed, while the incident angle and the azimuthal angle can be changed. They define $\delta$, the angle between the sample \emph{c}-axis and the transferred momentum \emph{$\textbf{q}$} (red arrow). The projection of \emph{$\textbf{q}$} onto the sample \emph{ab}-plane, \emph{$\textbf{q}_{\parallel}$}, is also shown. In the experiment $\delta$ is changed by rotating the sample around a vertical axis. In this way the regions of the 2D reciprocal space indicated by thick lines can be covered (panel a, to the right). The [1,0] and [1,1] directions correspond to $\phi_{i} = 0$ and $45^\circ$ respectively. Panel c: the orientation of the atomic spin moments is also a parameter in the cross section calculations.}
  \label{Fig01}
\end{indented}
\end{figure}

LCO, SCOC and NdBCO measurements were performed at the ADRESS beamline~\cite{StrocovJSR2010} of the Swiss Light Source at the Paul Scherrer Institute (SLS-PSI) in Villigen (CH) using the SAXES~\cite{SAXES} spectrometer. At the Cu $L_3$ edge (approximately 930 eV) the combined energy resolution is $\Delta E=130$ meV FWHM with a data point sampling of 14.1 meV/data-point. CCO RIXS spectra were recorded at the ID08 beamline of the European Synchrotron Radiation Facility (ESRF) in Grenoble (FR) using the AXES~\cite{AXES,DiNardo} spectrometer. The combined energy resolution here was $\Delta E=240$ meV FWHM (31.0 meV/data point) at the same incident energy. In both cases, the incident photon energy was finely tuned at the $L_3$ peak by inspection of the absorption spectrum. The linear polarization of the incident beam could be set either perpendicular (vertical polarization, $\sigma$, hereafter) or parallel (horizontal, $\pi$) to the scattering plane. The polarization of the outgoing beam was not detected. Spectra were obtained as the sum of 6 to 12 partial spectra of 5 or 10 minutes each. Off resonance elastic peaks were periodically measured on a graphite powder to determine the zero energy loss of the spectra. The pressure in the measurement vacuum chamber was better than $3\times10^{-9}$ mbar. All the measurements were done at 15 K (except LCO that was measured at room temperature).

In figure~\ref{Fig01} the scattering geometry is shown. The beam hits the sample surface at incident angles $\theta_{i}$ and $\phi_{i}$ and the outgoing beam is collected at angles $\theta_{o}$ and $\phi_{o}$ in the \emph{xyz} reference system; in general the scattering angle 2$\theta$ is determined by $\theta_{i}$, $\phi_{i}$, $\theta_{o}$ and $\phi_{o}$, but in our case $\phi_{o} = \phi_{i}$ so that 2$\theta$ depends only on $\theta_{i}$ and $\theta_{o}$. The momentum \emph{$\textbf{q}$} transferred to the sample due to the scattering process is shown by the red arrow. Its projection onto the sample \emph{ab}-plane ($\textbf{\emph{q}}_{\parallel}$) is also shown since the meaningful reciprocal space is usually 2D. The scattering angle 2$\theta$ is fixed at 130$^{\circ}$ for the AXES spectrometer, while it could be set either at 90$^{\circ}$ or 130$^{\circ}$ for SAXES, in all cases lying in the horizontal plane. It should be noted that, for a fixed 2$\theta$, \emph{$\textbf{q}$} is fixed, but \emph{$\textbf{q}_{\parallel}$} can be easily changed both in magnitude and direction in the sample 2D reciprocal space by rotating the sample itself around an axis perpendicular or parallel to the scattering plane. For the sake of simplicity we introduce the angle $\delta$ between the transferred momentum \emph{$\textbf{q}$} and the sample \emph{c}-axis, laying in the scattering plane. With this notation, $\left|\textbf{\emph{q}}_{\parallel}\right|=2\left|\emph{\textbf{k}}\right|\sin(\theta)\sin(\delta)$, where $\textbf{\emph{k}}=\textbf{\emph{k}}_{\mathrm{i}}$ ($\approx\textbf{\emph{k}}_{\mathrm{o}}$) is the momentum carried by ingoing (outgoing) photons. For example, when $\delta=0$ (specular geometry), \emph{$\textbf{q}$} is parallel to the sample \emph{c}-axis and \emph{$\textbf{q}_{\parallel}$}=0 thus allowing us to probe excitations in the center of the BZ. On the other hand, if one could go at infinitely grazing incidence (emission), then $\delta=-\theta$ ($+\theta$) and the maximum magnitude of \emph{$\textbf{q}_{\parallel}$} would then be reached. The value of \emph{$\textbf{q}$} depends on the scattering angle: at $\hbar\omega_i = 930$~eV $\textbf{q}_{\parallel max}= 0.47$ {\AA}$^{-1}$ for 2$\theta=90^{\circ}$, and 0.77 {\AA}$^{-1}$ for 2$\theta=130^{\circ}$, respectively. All the measurements were performed at a fixed scattering angle 2$\theta$ and changing $\delta$ by rotating the sample around an axis vertical in the laboratory space and perpendicular to the scattering plane, in steps of 5$^{\circ}$ typically. Samples were aligned so to have \emph{$\textbf{q}_{\parallel}$} parallel to the [10] direction ($\phi_{i}=0$) or the [11] direction ($\phi_{i}=45^{\circ}$) in the BZ.

LCO and CCO 100 nm thick films were grown by pulsed laser deposition on (001) SrTiO$_3$ single crystals. Excimer laser charged with KrF ($\lambda$ = 248 nm, 25 ns pulse width) was used at laser fluence at about 2 J/cm$^2$ at the target. Growth temperatures were 800 $^{\circ}$C for LCO and 650 $^{\circ}$C for CCO. A partial oxygen pressure of about 0.1 mbar was employed to correctly oxidize the films during deposition. In the case of LCO the quality of the films was also in-situ controlled by reflection high-energy electron diffraction (RHEED) technique. The SCOC sample were laminar-like single crystals obtained by slow cooling in air of a melt of Sr$_2$CuO$_2$Cl$_2$ powder. X-ray powder diffraction (XRD) analysis was conducted on a Rigaku x-ray diffractometer with Cu K$\alpha$ radiation ($\lambda$ = 1.5418 {\AA}) and the electron microprobe was used for chemical analysis . Nd$_{1.2}$Ba$_{1.8}$CuO$_{6+x}$ ($x<0.1$) 100 nm thin film were deposited on SrTiO$_3$ (100) single crystals by diode high-pressure oxygen sputtering. Undoped NdBCO thin films were obtained by reducing the oxygen content by annealing as grown Nd$_{1.2}$Ba$_{1.8}$CuO$_{7-\delta}$ samples in Argon atmosphere (10 mbar) for 24 hours. The structural and morphological properties of the sample have been checked by x-ray diffraction, also using synchrotron radiation, and by atomic force microscopy~\cite{SalluzzoPRB2005}.  The samples are tetragonal and perfectly matched with the STO lattice ( $a$=$b$=3.905 {\AA}) while the $c$-axis is 11.81 {\AA}.

For LCO, CCO, NdBCO and SCOC the fraction of the BZ that could be explored along the [10] direction was 93.1\%, 94.4\%, 95.6\% and 97.3\%, respectively, for 2$\theta=130^{\circ}$. In the case of LCO and SCOC, it decreased at 56.9 (59.4)\% with 2$\theta$ set at 90$^{\circ}$. Moreover in the case of SCOC, measurements along the [11] direction were taken up to 68.8\% of the nuclear BZ boundary with 2$\theta=130^{\circ}$, i.e., the whole magnetic BZ could be spanned.

\section{Experimental spectra and cross section calculations}

\subsection{Overview of experimental spectra}

\begin{figure}
\begin{indented}
  \item{}\includegraphics[width=10cm]{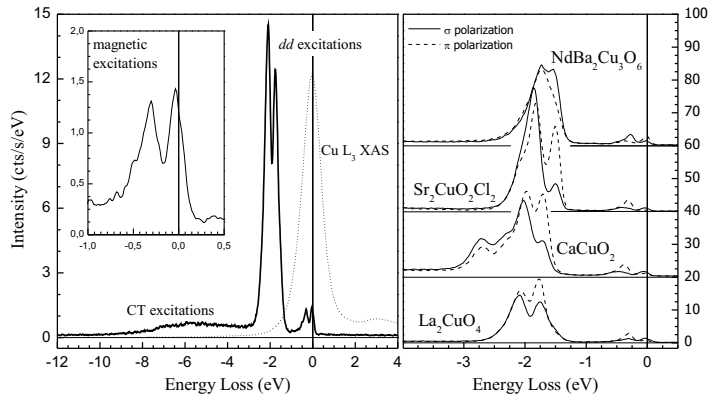}
  \caption{(Left panel) Example of Cu $L_3$ absorption (dashed) and RIXS (solid) spectra of LCO with $\sigma$ polarization. For RIXS the angles were set to $2\theta =130^{\circ}$, $\delta=+45^{\circ}$ (i.e.~20$^{\circ}$ grazing emission) and $\phi_{i}=0$. One can immediately recognize charge transfer (CT), \emph{dd} and magnetic excitations at different energy ranges. A closer look of the mid-infrared energy region is given in the inset. (Right panel) RIXS spectra for LCO, SCOC, CCO and NdBCO in the same experimental geometry as left panel.}
  \label{Fig02}
\end{indented}
\end{figure}

In the left panel of figure~\ref{Fig02} a Cu $L_3$ edge RIXS spectrum with $\sigma$ polarization of LCO is shown. The scattering angle 2$\theta$ was set at 130$^{\circ}$ and the sample was rotated at $\delta=+45^{\circ}$, i.e., 20$^{\circ}$ grazing emission, and $\phi_{i}=0$ (\emph{$\textbf{q}_{\parallel}$} parallel to the [10] direction in reciprocal space). The very high resolution allows us to clearly recognize different excitations depending on the energy scale. The spectrum is dominated by \emph{dd} excitations, which will be extensively discussed below. Both at higher and lower energies other features are found with a much (approximately one order of magnitude) lower counting rate. Excitations in the mid-infrared region (up to $\sim$500~meV) are expanded in the inset and are known to have a magnetic character: they are a combination of a dispersing magnon, a continuum given by bimagnons and other multi-magnon excitations, phonon peaks and an elastic zero-loss line~\cite{BraicovichPRL2009,LucioPhaseSep,GrioniPRLSCOC}. At higher energies CT excitations are also visible, as a broad distribution, via the Cu 3\emph{d}-O 2\emph{p} hybridization. By looking at the absorption spectrum (dashed line), it can also be noted that features above 1~eV energy loss, such as \emph{dd} excitations, are weakly affected by self-absorption because of the reduced absorption coefficient 2~eV below the resonance peak. In the right panel of figure~\ref{Fig02}, \emph{dd} excitations of LCO are compared to those of SCOC, CCO and NdBCO both for $\sigma$ (solid) and $\pi$ (dashed line) polarization. The effect of changing incident photon polarization is considerable. The spectral shape changes drastically for the four compounds, reflecting differences in the Cu$^{2+}$ coordination. In order to recognize the symmetry of the RIXS final state we carried out systematic measurements with both polarization as a function of $\delta$ (or $\theta_{i}$, accordingly) for a fixed 2$\theta$ and fitted the spectra to obtain an estimate of the $dd$ excitations energies. The fitting procedure is presented in the following sub-section and is based on single ion model calculations of the RIXS cross sections.

\subsection{Single ion model cross section calculations}

\begin{figure}
\includegraphics[height=.8\textheight]{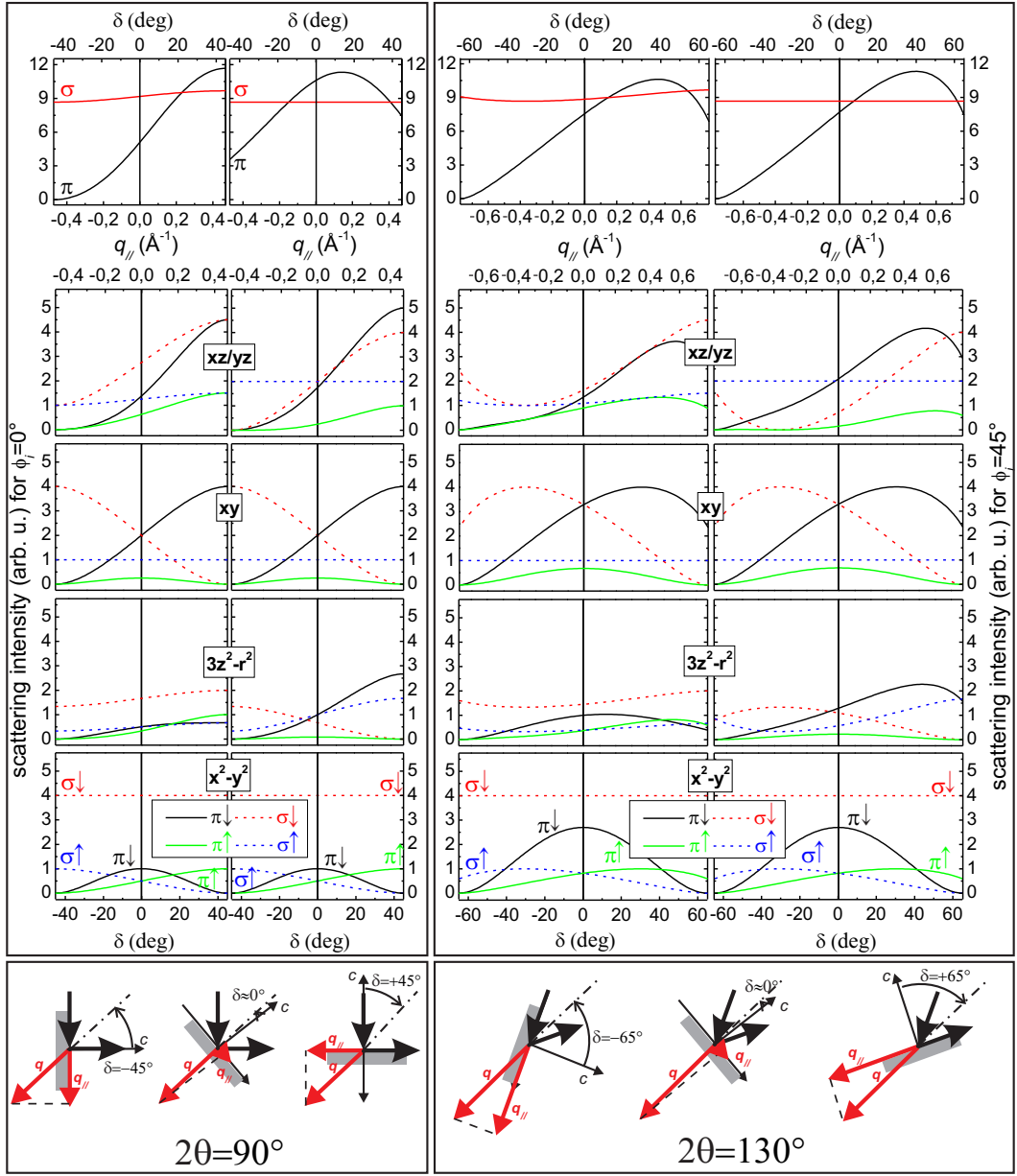}
  \caption{ Single ion Cu$^{2+}$ $L_3$ edge RIXS cross section for 2$\theta=90^{\circ}$ (left panel) and 2$\theta=130^{\circ}$ (right panel) and for two different angles $\phi_{i}=0$ and $\phi_{i}=45^{\circ}$ both for $\sigma$ and $\pi$ polarization. The bottom panels show sketches of the experimental geometry. The spin is fixed in the calculation along the [110] direction ($\theta_{s}=90^{\circ},\phi_{s}=45^{\circ}$). All the possible \emph{dd} excitations including the elastic and spin-flip final states are considered. As already explained in figure~\ref{Fig01} to a given $\delta$ (bottom axis scale) corresponds an in-plane transferred momentum $\textbf{\emph{q}}_{\parallel}$ (top axis scale). In the top panels the sums of all \emph{dd} excitations (excluding elastic and spin-flip) cross sections are given.}
  \label{Fig03}
\end{figure}

In cuprates, Cu ions are known to be mostly in the Cu$^{2+}$ oxidation state~\cite{Ghijsen1990} corresponding to a 3$d^9$ electronic configuration. In the crudest approximation, we only consider the atomic states of a Cu$^{2+}$ ion. RIXS is a second order process and is described by the Kramers-Heisenberg (KH) equation. The scattering is here modeled in two steps: first one 2$p_{3/2}$ electron is resonantly promoted into the 3$d$ states by absorption of a photon (intermediate state). Because of the large 2$p$ spin-orbit splitting ($\sim20$~eV), interference effects with the 2$p_{1/2}$ states can be neglected. The only available intermediate state is given by a fully occupied 3$d$ shell and one hole in the four-fold degenerate 2$p_{3/2}$ core level. The second step is the radiative deexcitation of one 3$d$ electron into the 2$p_{3/2}$ levels. In a short notation, the process can be written as the sequence 2$p^4_{3/2}3d^9\rightarrow2p^3_{3/2}3d^{10}\rightarrow2p^4_{3/2}3d^{9*}$, where the $^*$ indicates that the final states can be either the ground state or an excited state with the 3$d$ hole occupying a state with different orbital symmetry and/or spin.

Matrix elements entering the KH equation for the calculation of the RIXS cross sections are here calculated in the atomic approximation as follows. We consider the scattering ion as a hydrogen-like system of a single positively charged particle (hole) and we express the angular part of the one-particle wave functions ($p$ and $d$ orbitals) in spherical harmonics. The radial part of the integrals is the same for all the considered transitions and is thus neglected also in the expressions of the $2p$ and $3d$ states. The local symmetry is D$_{4h}$, so the commonly used set of atomic $d$ orbitals are well suited to represent the atomic states in terms of symmetry:
\begin{eqnarray}
    d_{x^2-y^2}&=&\frac{1}{\sqrt{2}}\(Y_{22}+Y_{2\bar{2}}\)\\
    d_{3z^2-r^2}&=&Y_{20}\\
    d_{xy}&=&-\frac{i}{\sqrt{2}}\(Y_{22}-Y_{2\bar{2}}\)\\
    d_{yz}&=&-\frac{i}{\sqrt{2}}\(Y_{21}+Y_{2\bar{1}}\)\\
    d_{xz}&=&\frac{1}{\sqrt{2}}\(Y_{21}-Y_{2\bar{1}}\)
\end{eqnarray}
The latter two are degenerate for D$_{4h}$ symmetry. As in all layered cuprates the four in-plane O ions are the Cu nearest neighbors, the unique 3$d$ hole present in the ground state has an almost pure  ${x^2-y^2}$ character~\cite{CTChenPRL1992}.  Thus the \emph{dd} excitations can be described as the transfer of the 3$d$ hole from the $d_{x^2-y^2}$ orbital to the $d_{\mathrm{3z^2-r^2}}$, $d_{xy}$, $d_{yz}$ or $d_{xz}$ orbitals. Also the case of $d_{x^2-y^2}$ final state can be viewed as a $dd$ excitation, giving rise to an elastic scattering or to a pure spin-flip excitation depending on the final spin state.

Superimposed to the CF splitting of 3$d$ orbitals is the additional spin splitting of each state due to super-exchange interaction with neighboring in-plane Cu$^{2+}$ ions (spin orbit of 3$d$ states is neglected) which doubles the number of possible final states. In this work we will not consider the low energy part of the spectra because the description of the set of possible final states cannot be done in a single ion model. In fact, the local spin-flip is not an eigenstate of the 2D Heisenberg AFM lattice: spin waves should then be introduced to account for the dispersion of pure magnetic excitations, as already demonstrated in references~\cite{AmentPRL2009,LucioPhaseSep}.

Although we do not look at pure magnetic excitations, in order to correctly calculate the RIXS cross sections the local atomic spin orientation has to be explicitly taken into account. As the $3d$ spin-orbit interaction is weak in Cu~\cite{GhiringhelliPRB2002} spin is a good quantum number for both the ground and the final states. We use here the Pauli matrices $\widetilde{\sigma}_x, \widetilde{\sigma}_y, \widetilde{\sigma}_z$. The arbitrary spin orientation of the initial hole is defined by the angles $\theta_{s}$ and $\phi_{s}$ and the eigenvectors of
\begin{equation}
    \widetilde{\sigma}(\theta_{s},\phi_{s})=\(\widetilde{\sigma}_x\cos\phi_{s}+\widetilde{\sigma}_y\sin\phi_{s}\)\sin\theta_{s}+\widetilde{\sigma}_z\cos\theta_{s}
\end{equation}
give the weights for spin up and spin down components along $z$ for the generic spin direction. For instance, if we assume the ground state spin direction to be ``down'' in the hole representation, the ground state wave function and the spin-flip wave function are written as
\begin{eqnarray}
    d_{x^2-y^2}^{\downarrow}&=&\frac{1}{\sqrt{2}}\[U_-\(Y_{22}^{\uparrow}+Y_{2\bar{2}}^{\uparrow}\)+D_-\(Y_{22}^{\downarrow}+Y_{2\bar{2}}^{\downarrow}\)\]\\
    d_{x^2-y^2}^{\uparrow}&=&\frac{1}{\sqrt{2}}\[U_+\(Y_{22}^{\uparrow}+Y_{2\bar{2}}^{\uparrow}\)+D_+\(Y_{22}^{\downarrow}+Y_{2\bar{2}}^{\downarrow}\)\]
\end{eqnarray}
where $U_-(U_+)$ and $D_-(D_+)$ are the components of the eigenvector corresponding to the negative (positive) eigenvalue of $\widetilde{\sigma}(\theta_{s},\phi_{s})$. The same holds for the other 3$d$ orbitals which, together with the ground state itself and the spin-flip state, represent the 10 different possible final states available for the transition, i.e $d_{x^2-y^2}^{\downarrow}, d_{x^2-y^2}^{\uparrow}, d_{3z^2-r^2}^{\downarrow}, d_{3z^2-r^2}^{\uparrow}, d_{xy}^{\downarrow}, d_{xy}^{\uparrow}, d_{xz}^{\downarrow}, d_{xz}^{\uparrow}, d_{yz}^{\downarrow}$ and $d_{yz}^{\uparrow}$.

On the contrary, 2\emph{p} states are strongly spin orbit coupled and spin is not a good quantum number; the energetically degenerate 2\emph{p}$_{3/2}$ orbitals are written in terms of spherical harmonics as follow:
\begin{eqnarray}
  p_{\frac{3}{2},-\frac{3}{2}}&=&Y_{1\bar{1}}^{\downarrow}\\
  p_{\frac{3}{2},-\frac{1}{2}}&=&\sqrt{\frac{1}{3}}Y_{1\bar{1}}^{\uparrow}+\sqrt{\frac{2}{3}}Y_{10}^{\downarrow}\\
  p_{\frac{3}{2},\frac{1}{2}}&=&\sqrt{\frac{2}{3}}Y_{10}^{\uparrow}+\sqrt{\frac{1}{3}}Y_{11}^{\downarrow}\\
  p_{\frac{3}{2},\frac{3}{2}}&=&Y_{11}^{\uparrow}
\end{eqnarray}
\\
In the approximation that the lifetimes of the 2\emph{p}$_{3/2}$ states are the same, the RIXS transition from the ground state $d_{x^2-y^2}^{\downarrow}$ to a given final state, say $d_{3z^2-r^2}^{\downarrow(\uparrow)}$, is given by
\begin{eqnarray}
    \sigma_{3z^2-r^2}^{\downarrow (\uparrow)} \propto \sum_{q'}\left|\sum_m
    \langle d_{3z^2-r^2}^{\downarrow(\uparrow)}|T^{\dagger}_{q'}| p_{\frac{3}{2},m}\rangle
    \langle p_{\frac{3}{2},m}|T_q|d_{x^2-y^2}^{\downarrow}\rangle
    \right|^2
\end{eqnarray}
where $m=-\frac{3}{2},-\frac{1}{2},\frac{1}{2},\frac{3}{2}$ runs over the intermediate states. $T_q=\sqrt{4\pi/(2q+1)}Y_{1q}(\theta,\phi)$ is the expression for the electric dipole operator in spherical harmonics with $q=-1,0,1$ for \emph{left}, \emph{linear} ($z$) and \emph{right} polarized light, respectively. The sum over $q'$ is required since we do not measure the polarization of the outgoing photons in our experiments. Matrix elements are thus simple integrals of three spherical harmonics which can be readily calculated~\cite{deGroot}.

Using this method we can calculate the scattering cross section for all the \emph{dd} excitations of cuprates for any incident and scattered directions and polarization of the photons and any atomic spin direction. For example, if the atomic spin is oriented along the [110] direction ($\theta_{s}=90^{\circ}$ and $\phi_{s}=45^{\circ}$) and 2$\theta=90^{\circ}$, the cross sections for the elastic and pure spin-flip transitions are given by the simple formulas
\begin{eqnarray}\label{HandCrosSectV}
    \sigma_{x^2-y^2,\sigma}^{\downarrow} &\propto& 4\\
    \sigma_{x^2-y^2,\sigma}^{\uparrow} &\propto& \sin^2\theta_{i}
\end{eqnarray}
in the case of incident $\sigma$ polarization and
\begin{eqnarray}\label{HandCrosSectH}
    \sigma_{x^2-y^2,\pi}^{\downarrow} &\propto& \sin^22\theta_{i}\\
    \sigma_{x^2-y^2,\pi}^{\uparrow} &\propto& \cos^2\theta_{i}
\end{eqnarray}
in the case of incident $\pi$ polarization and the scattering plane perpendicular to the sample surface.

In figure~\ref{Fig03}, the calculated cross sections are shown in the two particular cases $\phi_{i}=0$ and $\phi_{i}=45^{\circ}$ as a function of $\theta_{i}$ or alternatively of the in-plane transferred momentum \emph{$\textbf{q}_{\parallel}$} for 2$\theta=90^{\circ}$ (left panel) and 2$\theta=130^{\circ}$ (right panel). We underline that, contrarily to what previously stated in the literature~\cite{deGroot,vanVeenendaal}, pure spin-flip transitions are allowed as long as the spin is not parallel to the [001] direction~\cite{AmentPRL2009} (see the curves labeled $\sigma \uparrow$ and $\pi \uparrow$ for the $x^2-y^2$ final state in figure~\ref{Fig03}). For layered cuprates, this selection rule has the important consequence that single magnons can contribute to the Cu $L_3$ edge RIXS spectrum~\cite{AmentPRL2009} since here spins are known to lie always in the \emph{ab}-plane~\cite{VakninPRL1987,VakninPRB1989,VakninPRB1990}. We highlight the fact that the orientation of the spin within the $ab$ plane has no influence on the spin-flip cross section. This makes it simpler to measure the magnon dispersion in samples with multiple magnetic domains. However, some of the $dd$ cross sections do depend on the $\phi _s$ value (having fixed $\theta _s = 90^\circ$). By this dependence one could think of using the $dd$ excitation spectrum to determine the in-plane orientation of the spin in layered cuprates.

Calculations are then used to determine the energy and symmetry of Cu-3$d$ states by fitting the experimental data. It is important to note that cross sections are here calculated on the basis of symmetry properties of the angular part of the ground state and final state wave functions. By including the Cu 3$d$-O 2$p$ hybridization the symmetry of the problem would not change~\cite{Eskes,KotaniEPJB2005} and the dependence on the scattering geometry and photon polarization would remain the one calculated here, as it depends solely on symmetry properties~\cite{KotaniEPJB2005}.

Once the cross sections for all the possible final states are calculated, one has to construct the simulated spectra as a function of the energy loss $(\hbar \omega_{o}-\hbar \omega_{i})$ as
\begin{eqnarray}
\lefteqn{I(\hbar \omega_{o}-\hbar \omega_{i})=}\\
\nonumber        &&=\frac{1}{\pi}\sum_f \[ \frac{\Gamma_f\sigma_f^{\downarrow}}{\(\hbar \omega_{o}-\hbar \omega_{i}+E_f\)^2+\Gamma_f^2}+
    \frac{\Gamma_f\sigma_f^{\uparrow}}{\(\hbar \omega_{o}-\hbar \omega_{i}+E_f+2J_f\)^2+\Gamma_f^2}\]
\end{eqnarray}
where $f$ runs over ${x^2-y^2}$, $3z^2-r^2$, $xy$, $xz$ and $yz$. We assume here that the Lorentzian lifetime broadening $\Gamma_f$ is equal for states with the same orbital symmetry while the super-exchange coupling ($J_f$) is orbital dependent. From overlap considerations~\cite{Harrison} one finds that $J_{3z^2-r^2}=J_{x^2-y^2}/6=J/6$ and $J_{xy}=J_{xz}=J_{yz}=0$. $J_f$ gives the energy separation between a peak and its spin-split satellite: according to the 2D Ising model, in the ground state the energetic cost to flip one spin is $2J$, while in the excited states the exchange constant is reduced because of the smaller overlap with nearest neighbor orbitals. In the present work we have fixed $J = 130$ meV for all samples for simplicity, although it is known that its value is different ($\pm20\%$) from sample to sample. Indeed this difference would little affect our calculated spectra, due to the strong reduction or cancelation of $J_f$ with respect to $J$ as explained above.

\begin{figure}
\begin{indented}
  \item{}\includegraphics[height=8cm]{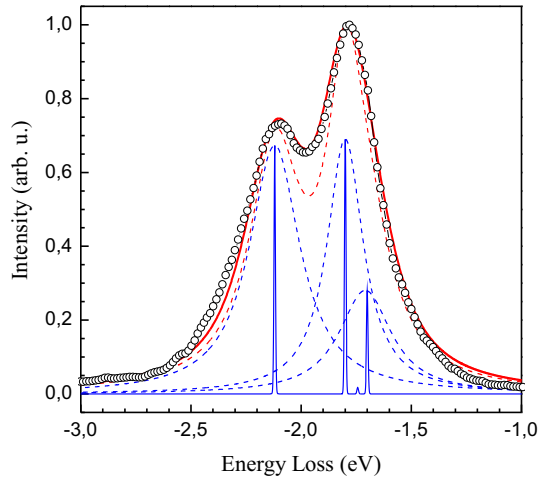}
  \caption{ Cu $L_3$ edge RIXS spectrum (open circles) of LCO taken at $\delta=0$, 2$\theta=90^{\circ}$ and $\sigma$ polarization. The theoretical spectrum is built by convoluting delta functions proportional to the calculated cross sections $\sigma_f$ (thick blue) with Lorentians to include finite lifetime broadening (dashed blue). Their sum gives the hypothetical spectrum measured with infinite resolution (dashed red). The spectrum to be compared to the experimental results is obtained after convolution with a Gaussian curve (solid red). The energy of the various $dd$ excitations ($E_f$) and the Lorentian widths ($\Gamma_f$) are used as adjustable parameters in the fitting procedure.}
  \label{Fig04}
  \end{indented}
\end{figure}

Finally $E_f$ is the energy of the final state with a given symmetry ($f$) and spin down ($\downarrow$). Geometry and polarization dependencies that enter the simulations through $\sigma_f$ are obviously fixed by the experiment. This imposes a severe constraint to the number of free fitting parameters which is thus limited to the energy positions of the 3\emph{d} states (namely $E_{3z^2-r^2}$, $E_{xy}$ and $E_{xz/yz}$, while $E_{x^2-y^2}=0$ by definition) and to their Lorentzian lifetime broadening ($\Gamma_{3z^2-r^2}$, $\Gamma_{xy}$ and $\Gamma_{xz/yz}$). The energies ($E_f$) of the final states should be considered as effective energies, thus taking into account both the ionic and the covalent part of the bond. Figure~\ref{Fig04} shows an example of the fitting procedure for LCO ($\delta=0$, 2$\theta=90^{\circ}$, $\sigma$ polarization). Thick blue lines are $\delta$-like functions whose heights are proportional to the calculated cross sections ($\sigma_f$); they are convoluted with Lorentzian functions to take into account the lifetime broadening of the final states ($\Gamma_f$) to obtain the dashed curves. Their sum gives the red dashed line which is eventually convoluted with a Gaussian (whose FWHM matched to the effective experimental resolution) to obtain the simulated Cu $L_3$ edge RIXS spectrum (red line).

\section{Determination of $dd$ excitation energies and discussion}

\subsection{Fitting results}

\begin{figure}
\begin{indented}
  \item{} \includegraphics[width=10cm]{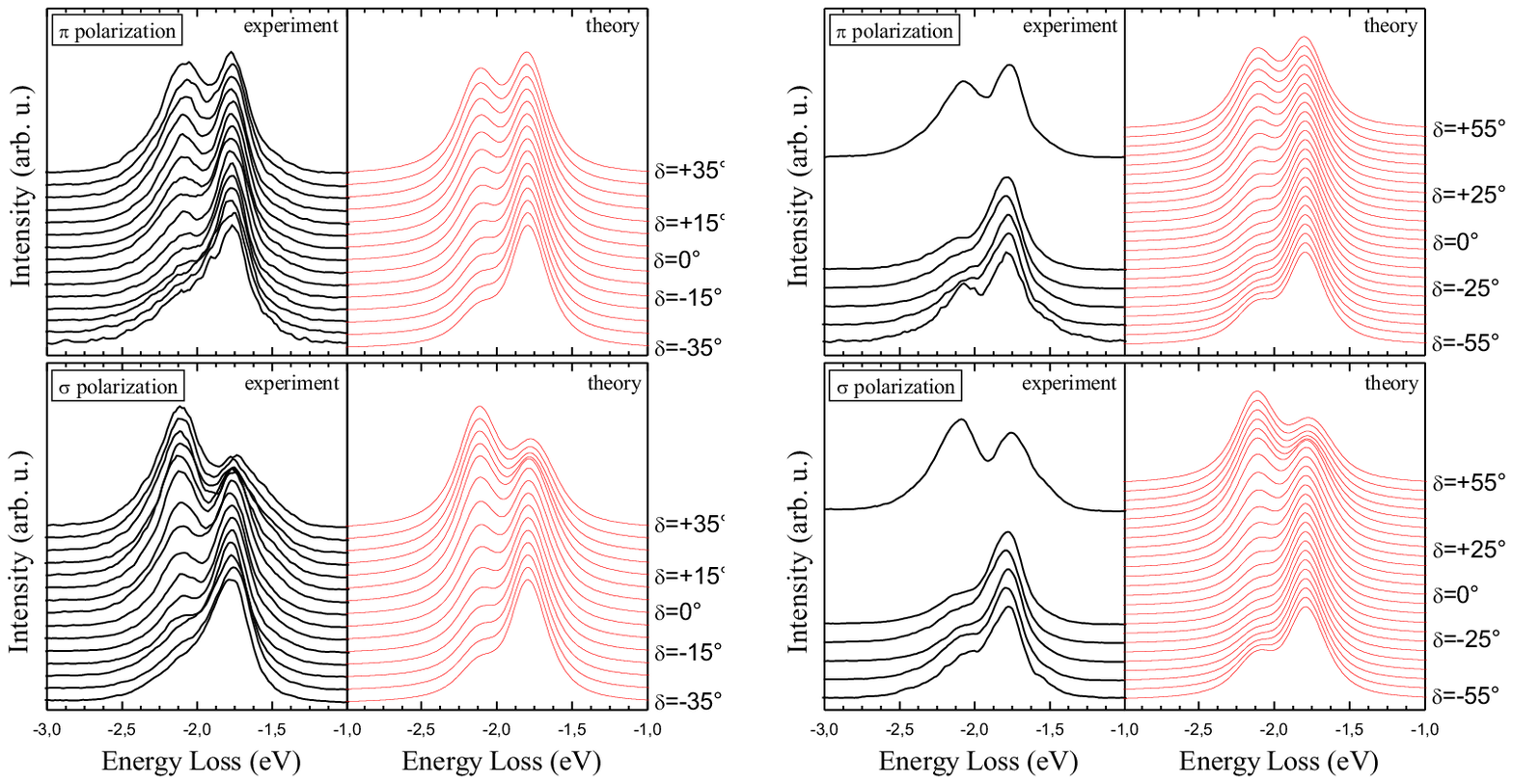}
  \caption{ Measured (black line) and calculated (red line) RIXS spectra of LCO for various $\delta$, ranging from 10$^{\circ}$ grazing incidence to 10$^{\circ}$ grazing emission with respect to the sample surface in steps of 5$^{\circ}$. Dispersion along [10] direction: $\phi_{i}=0$; 2$\theta=90^{\circ}$ (left panel) and 2$\theta=130^{\circ}$ (right panel).}
  \label{Fig05}
\end{indented}
\end{figure}

\begin{figure}
\begin{indented}
  \item{}\includegraphics[width=10cm]{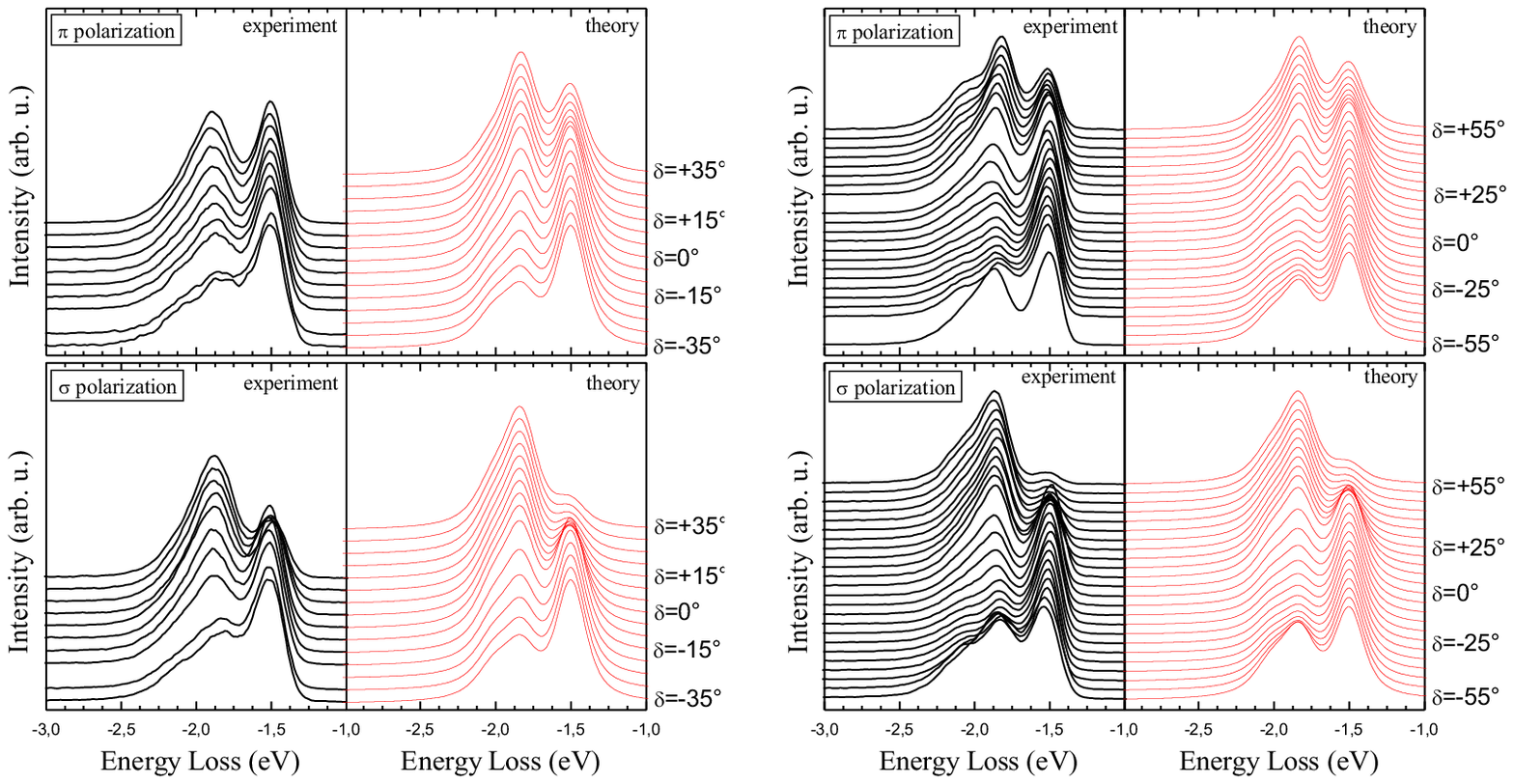}
  \caption{ Measured (black line) and calculated (red line) RIXS spectra of SCOC for various $\delta$, ranging from 10$^{\circ}$ grazing incidence to 10$^{\circ}$ grazing emission with respect to the sample surface in steps of 5$^{\circ}$. Dispersion along [10] direction: $\phi_{i}=0$; 2$\theta=90^{\circ}$ (left panel) and 2$\theta=130^{\circ}$ (right panel).}
  \label{Fig06}
\end{indented}
\end{figure}

\begin{figure}
\begin{indented}
  \item[]\includegraphics[height=9cm]{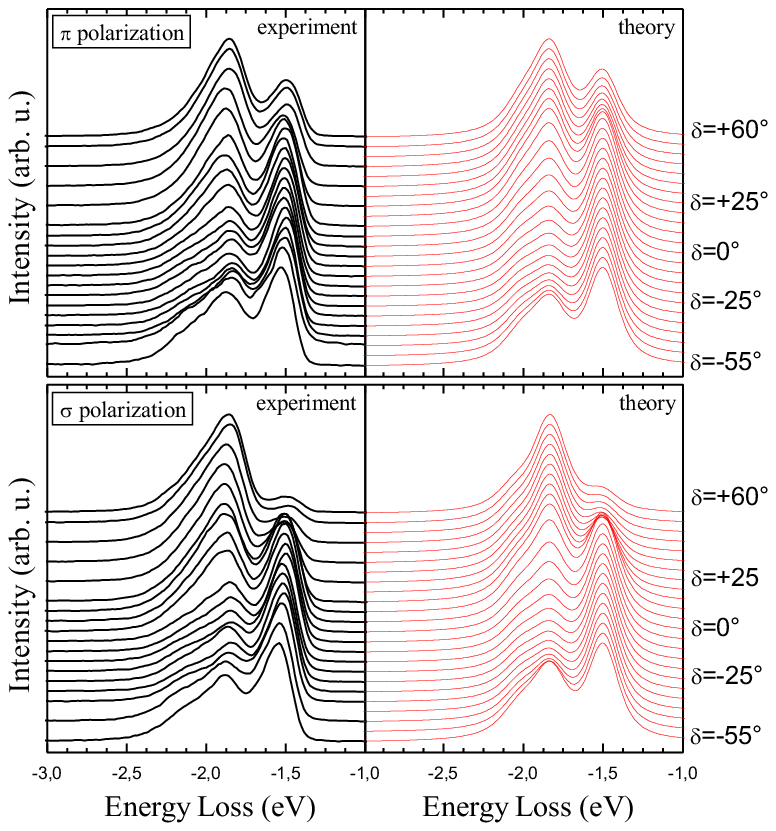}
  \caption{Measured (black line) and calculated (red line) RIXS spectra of SCOC for various $\delta$, ranging from 10$^{\circ}$ grazing incidence to 10$^{\circ}$ grazing emission with respect to the sample surface in steps of 5$^{\circ}$. Dispersion along [11] direction: $\phi_{i}=45^{\circ}$; 2$\theta=130^{\circ}$.}
  \label{Fig07}
\end{indented}
\end{figure}

\begin{figure}
\begin{indented}
  \item[]\includegraphics[height=9cm]{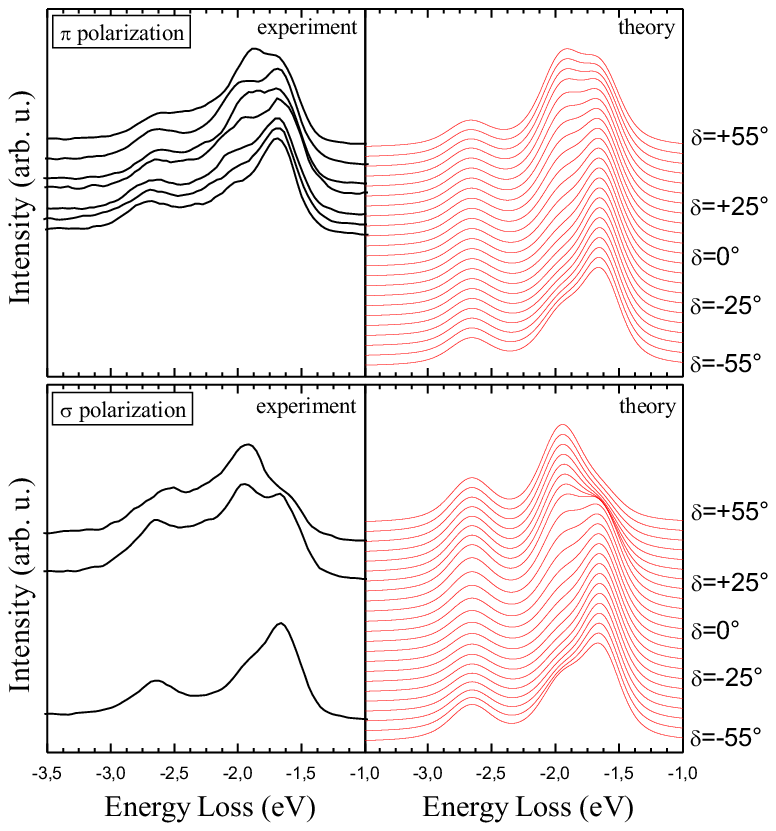}
  \caption{Measured (black line) and calculated (red line) RIXS spectra of CCO for various $\delta$, ranging from 10$^{\circ}$ grazing incidence to 10$^{\circ}$ grazing emission with respect to the sample surface in steps of 5$^{\circ}$. Dispersion along [10] direction: $\phi_{i}=0$; 2$\theta=130^{\circ}$.}
  \label{Fig08}
\end{indented}
\end{figure}

\begin{figure}
\begin{indented}
  \item[]\includegraphics[height=9cm]{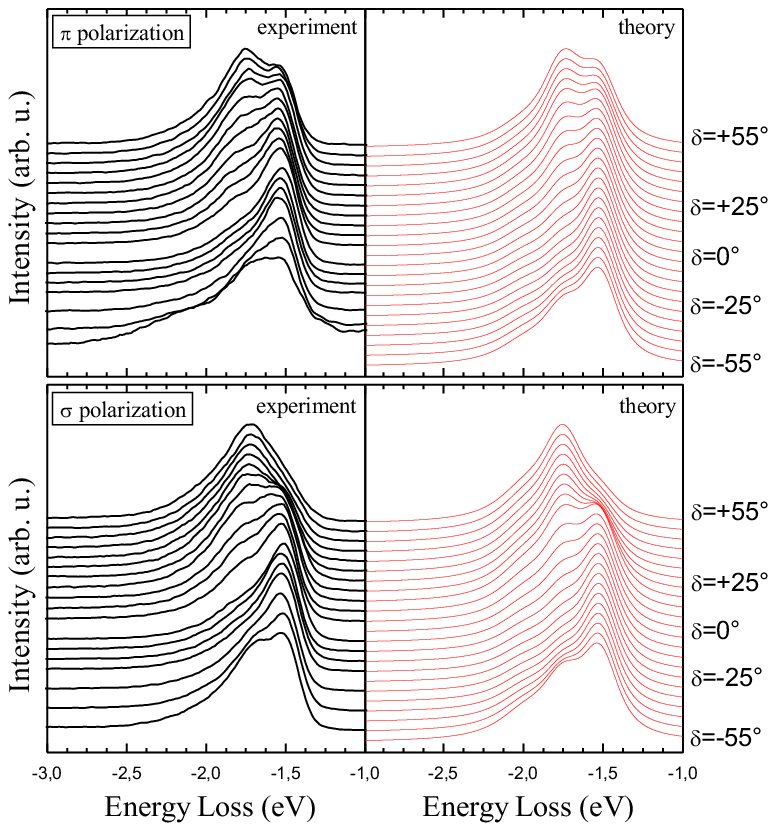}\\
  \caption{Measured (black line) and calculated (red line) RIXS spectra of NdBCO for various $\delta$, ranging from 10$^{\circ}$ grazing incidence to 10$^{\circ}$ grazing emission with respect to the sample surface in steps of 5$^{\circ}$. Dispersion along [10] direction: $\phi_{i}=0$; 2$\theta=130^{\circ}$.}
  \label{Fig09}
\end{indented}
\end{figure}

In this section, the experimental trends are shown together with the results of the simulations. In figure~\ref{Fig05}, measured and calculated RIXS spectra of LCO are compared, both for 2$\theta=90^{\circ}$ and 2$\theta=130^{\circ}$ with $\phi_{i}=0$. From the top spectrum to the bottom one $\delta$ is changed in steps of 5$^{\circ}$ from 10$^{\circ}$ grazing emission to 10$^{\circ}$ grazing incidence. As the incidence angle is changed, \emph{$\textbf{q}_{\parallel}$} varies accordingly. However, it is evident from the experimental results that the dispersion of \emph{dd} excitations, if any, is very small both for 2$\theta=90^{\circ}$ and 2$\theta=130^{\circ}$. This finding seems to be general and holds true also in the cases of SCOC (figures~\ref{Fig06} and \ref{Fig07}), CCO (figure\ref{Fig08}) and NdBCO (figure\ref{Fig09}). Within the present experimental accuracy the $dd$ excitations do not disperse in any of the undoped layered cuprates measured by us.

The fact that $dd$ excitations are fixed in energy when changing $\delta$ allowed us to fit all the spectra for a given sample using fixed values of $E_f$ and $\Gamma_f$. Using the redundance of the experimental basis the values of the free parameters $E_f$ and $\Gamma_f$ could thus be optimized. The high degree of confidence reached is made evident by the excellent agreement of the fitted spectra to the experimental ones shown in the figures. For clarity we  remind here the input parameters of our fitting. The RIXS cross sections $\sigma _f$ are calculated for each final state within the ionic model; the superexchange interaction is fixed to $J=130$ meV for all samples; the Gaussian broadening is set to 130 meV FWHM; the energy $E_f$ and Lorentzian broadening $\Gamma_f$ are the free parameters  for optimizing the fitting and are thus the result of the comparison of calculated and measured spectra.

The results for the four samples are summarized in Table \ref{TableI}. For LCO and SCOC we used data measured both at $2\theta = 90^\circ$, $130^\circ$ for $\phi_i=0$. For SCOC we employed data with $2\theta = 130^\circ$ and $\phi_i=45^\circ$. For CCO and NdBCO the used data were at $2\theta = 130^\circ$ and $\phi_i=0$ only. Although the spectra of CCO shown in figure~\ref{Fig08} were measured with lower resolving power ($\Delta E=240$~meV), the fitting results are perfectly compatible with the (few) spectra taken at higher resolution ($\Delta E=130$~meV, figure~\ref{Fig02}). However, here the better resolution makes evident an additional feature at about 2.4 eV, which could not be assigned within our model. Possibly the extra peak could be related to oxygen vacancies affecting the local environment of Cu ions. Independent results on magnetic excitations of the same sample~\cite{LucioStrain} seem to further support this interpretation.

Finally we notice that the fitting works well for NdBCO too, in spite of the presence in this sample of two different Cu sites. Namely in this YBCO-like system Cu ions can be found also outside the CuO$_2$ planes, in the so-called CuO chains (although in this undoped compound chains are actually broken). As in the simulations we assume that only one specie is contributing to the RIXS spectrum, we attribute this result to the extreme selectivity provided by the resonance in the excitation step.

\begin{table}
    \caption{\label{TableI}Parameters used in the calculations to fit the experimental data. Also listed are the values of the effective CF parameters and those for a pure covalent picture within a 2D cluster. In all except the LCO case, covalent parameters could not be defined (n.d.) as explained in the text. For all samples the superexchange $J$ was fixed to 130 meV.}

    \begin{tabular}{c|c|c|c|c|c}
        \br
        &La$_2$CuO$_4$ & Sr$_2$CuO$_2$Cl$_2$ & CaCuO$_2$ & Sr$_{0.5}$Ca$_{0.5}$CuO$_2$ &NdBa$_2$Cu$_3$O$_7$\\
        \mr

        $E_{3z^2-r^2}$ $\(\Gamma_{3z^2-r^2}\)$ [eV]&1.70 (.14)&1.97 (.10)&2.65 (.12)& 2.66  &1.98 (0.18)\\
        $E_{xy}$ $\(\Gamma_{xy}\)$ [eV]&1.80 (.10)&1.50 (.08)& 1.64 (.09)& 1.56 &1.52 (0.10)\\
        $E_{xz/yz}$ $\(\Gamma_{xz/yz}\)$ [eV]&2.12 (.14)&1.84 (.10)&1.95 (.12)& 1.93 &1.75 (0.12)\\

        $\Delta E_{e_g}$ &1.70&1.97&2.65&2.66&1.98\\
        $\Delta E_{t_{2g}}$ &0.32&0.33&0.31&0.36&0.23\\

        10$Dq$ [eV]&1.80 &1.50&1.64&1.56& 1.52\\
        $D_s$ [eV]&0.29&0.33&0.42& 0.43 & 0.32\\
        $D_t$ [eV]&0.11&0.13&0.19& 0.19 & 0.14\\

        $\Delta_{pd}$ [eV]& 2.20 & n.d. & n.d. & n.d. & n.d. \\
        $T_{pd}$ [eV]& 3.20 & n.d. & n.d. & n.d. & n.d. \\
        $T_{pp}$ [eV]& 0.81 & n.d. & n.d. & n.d. & n.d. \\

        \br
    \end{tabular}
\end{table}

\subsection{Discussion}

\begin{figure}[t]
\begin{indented}
  \item[] \includegraphics[width=10cm]{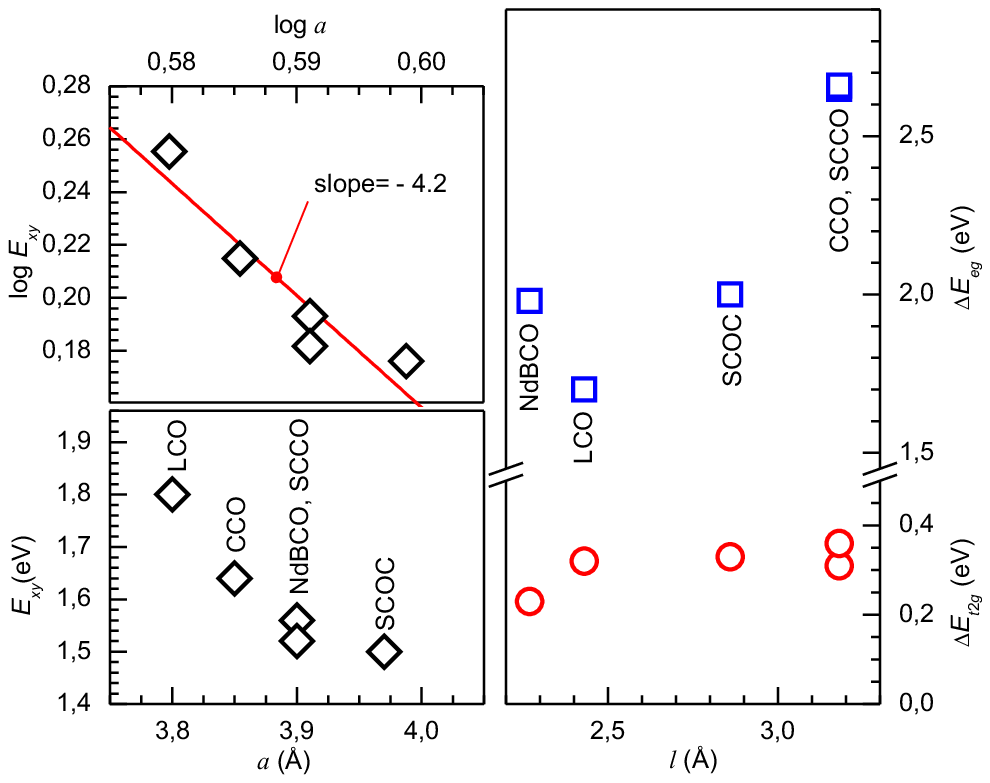}\\
  \caption{(Left panels) The effective CF parameter 10$Dq$ as a function of the lattice constant: linear (lower) and logarithmic (upper panel) scale. (Right panel) The splitting of the $t_{2g}$ and $e_g$ states as a function of the out-of-plane nearest neighbor distance $l$.}
  \label{Fig11}
\end{indented}
\end{figure}

The energies of the 3$d$ states obtained through our RIXS measurements are evidently related to the local coordination of Cu ions (symmetry and atomic distances). In fact, the 4 materials share the four-fold planar coordination that gives the famous CuO$_4$ plaquettes present in all cuprate superconductors. However, the in-plane Cu-O distances vary from a minimum in LCO to a maximum in SCOC: the lattice parameters are $a=3.80$~{\AA} for LCO~\cite{VakninPRB1990}, 3.85~{\AA} for CCO~\cite{QinPhysC2005}, 3.90~{\AA} for NdBCO~\cite{Petrikin2000} and 3.97~{\AA} for SCOC~\cite{VakninPRB1990}. The out-of-plane coordination is even more diverse. CCO has no apical ligands and, for that, it is often called infinite layer compound. Both LCO and SCOC have two apical ligands, symmetric with respect to the basal plane, at larger distance than the in-plane O (tetragonally distorted octahedral symmetry); in LCO apical O$^{2-}$ ions are at 2.43~{\AA} from Cu$^{2+}$, in SCOC Cl$^-$ ions are at 2.86~{\AA}. Finally NdBCO has a double layer YBCO-like structure, i.e., only one apical oxygen at 2.27~{\AA} from the CuO$_2$ plane.

The relation of the $dd$ excitation energies to the local structural properties of our samples is evident and relatively simple. Before comparing the results to any theoretical model we highlight these relations in a purely phenomenological way. The experimentally determined energies for each final state symmetry are listed in Table~\ref{TableI} and are referred to the $x^2-y^2$ ground state. A graphical presentation is made in figure~\ref{Fig11}. We use also the results of Sr$_{0.5}$Ca$_{0.5}$CuO$_2$ (SCCO, same structure as CCO but with larger in-plane lattice parameter $a=3.90$~{\AA}) are added, although the spectra are not shown here~\cite{LucioStrain}. The $xy$ excitation energy increases when decreasing $a$. When displayed in a logarithmic scale a power relation can be highlighted: a best fit to our data gives $E_{xy}\propto a^{-4.2}$. The splitting of the $t_{2g}$ states ($\Delta _{t_{2g}} = E_{xz/yz} - E_{xy}$) is rather small and almost independent of the out-of-plane lattice parameter $l$ and of the presence of apical ligands. On the contrary, the splitting of the $e_g$ states ($\Delta _{e_g} = E_{3z^2-r^2} - E_{x^2-y^2}$) varies considerably from sample to sample but a univocal trend vs the Cu-ligand distances ($a/2$ and $l$) cannot be found due to the other parameters at play (ligand element and valence, symmetry with respect to the basal plane). However, if we do not consider NdBCO, which has a pyramidal coordination, a qualitative trend emerges: the $e_g$ splitting incrases when the apical ligand is farther from the Cu ion. And the energy of the $3z^2-r^2$ state differs as much as almost 1 eV in LCO and CCO. This demonstrates that the local coordination can hugely impact on the $e_g$ splitting.

\begin{figure}[t]
\begin{indented}
  \item[]\includegraphics[width=7cm]{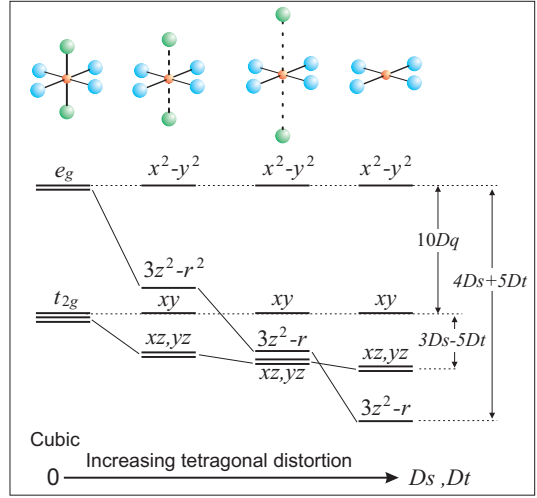}
  \caption{ Local crystal field and energy level diagram for \emph{d} orbitals in spherical, cubic (O$_h$) and tetragonal (D$_{4h}$) symmetry.}
  \label{Fig10}
\end{indented}
\end{figure}

The widely known crystal field (CF) model can be used to understand the trends in the experimental results. We take a single site, purely ionic picture, where the energies of the $3d$ orbitals is determined by the symmetry and strength of the non-central Coulomb field produced by the charged ligands nearby the central Cu$^{2+}$ ion. Beside an addictive constant, the energy levels of $d$ states in the tetragonally distorted octahedral symmetry are usually written as~\cite{Bersuker}
\begin{eqnarray}
  E_{x^2-y^2}^{\mathrm{CF}} &=& 6Dq+2D_s-D_t \\
  E_{3z^2-r^2}^{\mathrm{CF}} &=& 6Dq-2D_s-6D_t \\
  E_{xy}^{\mathrm{CF}} &=& -4Dq+2D_s-D_t\\
  E_{xz/yz}^{\mathrm{CF}} &=& -4Dq-D_s+4D_t
\end{eqnarray}
where $Dq$ (or 10$Dq$), $D_s$ and $D_t$ are the CF parameters depending on the geometrical arrangement of point charges around the Cu$^{2+}$ ion. In particular, 10$Dq$ gives the energy splitting between the $e_{g}$ and $t_{2g}$ orbitals, while $4D_s+5D_t$ and $3D_s-5D_t$ give the splitting of $e_{g}$ ($\Delta E_{e_g}$) and $t_{2g}$ ($\Delta E_{t_{2g}}$) orbitals, respectively (see figure~\ref{Fig10}). It can be shown~\cite{Bersuker} that within the CF model 10$Dq$ scales with a power law of the in-plane lattice parameter $a$, i.e., 10$Dq\sim a^{-n}$, with $n=5$. The value of the effective CF parameters 10$Dq$, $D_s$ and $D_t$ were determined by comparison with the experimental results and are given in Table~\ref{TableI}. The single ion CF model neglects completely the Cu-ligand orbital overlaps and some important physical properties are totally missing, such as the super-exchange interaction. However, it can capture the power law dependence of $E_{xy}$ on $a$, although the exponent $n$ is overestimated ($n=5$ instead of 4.2). On the other hand, although it is always possible to find a combination of $10Dq$, $D_s$ and $D_t$ compatible with the experimental results, it might happen that their values have little physical meaning. In particular $10Dq$, which is widely used in various calculations from single ion to cluster or impurity models, has to be regarded here as an \emph{effective} parameter, i.e., the value obtained in CF cannot be applied directly to other models.

On the other hand, one can tentatively try to include the effect of covalency with the help of a 2D-cluster calculations: Eskes \etal~\cite{Eskes} proposed a cluster-model calculation of the electronic structure of CuO in which they treat the $d-d$ Coulomb and exchange interaction within full atomic multiplet theory and use symmetry-dependent Cu 3$d$-O 2$p$ hybridization to describe photoelectron spectroscopic data. In the model, they consider a (CuO$_4$)$^{6-}$ cluster (plaquette), but neglect apical oxygens. In our case, this model seems is only to be partially justified: in fact, we can expect that the 2D-cluster, which does not include the apical ligangs, cannot fully account for the dependence of $\Delta _{e_g}$ and $\Delta _{t_{2g}}$ on the out-of-plane Cu-ligand distance $l$. Within this covalent picture, the energy levels of $d$ states are written as
\begin{eqnarray}
  E_{x^2-y^2}^{\mathrm{cov}} &=& \frac{1}{2} \(\Delta_{pd} - T_{pp}\) - \sqrt{T_{pd}^2 + \frac{1}{4} \(\Delta_{pd} - T_{pp}\)^2} \\
  E_{3z^2-r^2}^{\mathrm{cov}} &=& \frac{1}{2} \(\Delta_{pd} + T_{pp}\) - \sqrt{\(\frac{T_{pd}}{\sqrt{3}}\)^2 + \frac{1}{4} \(\Delta_{pd} + T_{pp}\)^2} \\
  E_{xy}^{\mathrm{cov}} &=& \frac{1}{2} \(\Delta_{pd} + T_{pp}\) - \sqrt{\(\frac{T_{pd}}{2}\)^2 + \frac{1}{4} \(\Delta_{pd} + T_{pp}\)^2} \\
  E_{xz/yz}^{\mathrm{cov}} &=& \frac{1}{2} \Delta_{pd} - \sqrt{\(\frac{T_{pd}}{2\sqrt{2}}\)^2 + \frac{1}{4} \Delta_{pd}^2}
\end{eqnarray}
where $\Delta_{pd}$ is the charge-transfer energy, $T_{pd}$ is the ground state Cu 3$d$-O 2$p$ hybridization energy and $T_{pp}$ is the nearest-neighbor O 2$p$-O 2$p$ hybridization energy. For LCO, the values of the parameters were obtained by taking the value of $\Delta_{pd}$ from independent experiments (2.2 eV~\cite{EllisPRB2008}) and by fulfilling two of the three equations for $T_{pd}$ and $T_{pp}$. The values which better reproduce the experimentally determined energies of Cu-3$d$ states in LCO are $T_{pd}$=3.20 eV and $T_{pp}$=0.81 eV. However, this 2D-cluster cannot describe the energy sequence found for SCOC, CCO and NdBCO. In fact, in those compounds $E_{3z^2-r^2}<E_{xy}$ whereas $E_{3z^2-r^2}^{\mathrm{cov}}<E_{xy}^{\mathrm{cov}}$ for any value of $\Delta_{pd}$, $T_{pd}$ and $T_{pp}$, as it can be seen from the equations above. It is anyhow interesting to notice that this model predicts a power law for the $xy$ to $x^2-y^2$ energy splitting with exponent $n=3.5$, dictated by the dependence of the overlap integrals on the Cu-O distance~\cite{Harrison}. This value is rather close to the experimental one.

\section{Conclusions}

Using Cu $L_3$ RIXS we have unequivocally determined the energy of $dd$ excitations (directly related to the energy of Cu-3$d$ states) in several layered cuprates, parent compounds of high $T_c$ superconductors. These values were largely unknown until now, because other experimental techniques could provide only partial or ambiguous results. For that reason the possible role of $dd$ excitations in superconductivity of cuprates has been often controversial. Three main outcomes can be extracted from the comparison of the quantitative results summarized in Table~\ref{TableI}.

i) The $dd$ excitation average energy is about 1.9 eV and, more importantly, the minimum lies above 1.4~eV. This finding resolves speculations made in the literature about a possible role of $3z^2-r^2$ states lying as close as 0.5 eV to the $x^2-y^2$ ground state. The absence of $dd$ excitations in the mid-infrared spectral region greatly supports the hypothesis that magnetic excitations (up to 250-300 meV) play a major role in Cooper pairing in cuprate superconductors, as one can obtain from the solution of Eliashberg equations \cite{LeTacon:unp}. However, $dd$ excitations can still have an importance for the full description of high $T_c$ superconductivity \cite{Little2007}. In particular the fact that $E_{3z^2-r^2}$ greatly varies from sample to sample, is potentially very  important in determining $T_c$ \cite{Sakakibara2010}.

ii) Apical ligands do have an influence on $dd$ excitations, but a quantitative trend cannot be predicted by simple models such as crystal field or 2D-clusters. As a consequence more sophisticated calculations \cite{Hozoi2007} are needed in order to reproduce the energy position of 3$d$ states in cuprates case by case, as determined experimentally.

iii) A simple relation of the $xy$ state energy to the in-plane lattice parameter exists. It is a power law ($E_{xy}\propto a^{-n}$) with exponent $n \simeq 4.2$, i.e., in between the prediction of crystal field model (purely electrostatic, $n=5$) and that of a ``covalent'' model following Harrison's overlap integrals ($n=3.5$). The dependence of overlap integrals on the interatomic distances is of great importance in cuprates as it drives the super-exchange interaction too, which provides the strong antiferromagnetic background in high $T_c$ superconductors (nevertheless one should not forget that overlap integrals are also strongly affected by Cu-O-Cu bond angles, a variable not considered in this work). Again, although indirectly, a better knowledge of $dd$ excitations can be of great help in a full description of superconductivity in cuprates.

In conclusion we have made a solid assessment of the $dd$ excitations in cuprates. These results can serve as experimental basis for advanced calculations of these same energies. Further measurements, both on parent compounds and on superconductors, could look for correlations between the $dd$ excitation spectrum and the superconducting properties of the cuprates.

\section{Acknowledgements}

Part of  this work was performed at the ADRESS beam line of the Swiss Light Source using the SAXES instrument jointly built by Paul Scherrer Institut (Switzerland), Politecnico di Milano (Italy) and EPFL (Switzerland). MMS, VB, LB and GG gratefully acknowledge M. W. Haverkort for his help in fixing bugs in the calculations and for enlightening discussions.

\end{document}